# The effects of resonant magnetic perturbations and charge-exchange reactions on fast ion confinement and neutron emission in the Mega Amp Spherical Tokamak

K G McClements[1], K Tani[2], R J Akers[1] Y Q Liu[1], K Shinohara[3], H Tsutsui[2] and S Tsuji-Iio[2]

[1] CCFE, Culham Science Centre, Abingdon, Oxfordshire, OX14 3DB, UK,
[2] Tokyo Institute of Technology, Ookayama Campus, 2-12-1 Ookayama, Meguro-ku, Tokyo 152-8550, Japan
[3] National Institutes for Quantum and Radiological Science and Technology, Naka Ibaraki 311-0193 Japan

E-mail: Ken.McClements@ukaea.uk

**Abstract**

The application of non-axisymmetric resonant magnetic perturbations (RMPs) to low current plasmas in the Mega Amp Spherical Tokamak (MAST) was found to be correlated with substantial drops in neutron production, suggesting a significant degradation of fast ion confinement. The effects of such perturbations on fast ions in MAST have been modelled using a revised version of a non-steady-state orbit-following Monte-Carlo code (NSS OFMC), in which the parametrization of fusion reaction rates has been updated and neutron rates have been recalculated. Losses of fast ions via charge-exchange (CX) with background neutrals and the subsequent reionization of fast neutrals due to collisions with bulk plasma particles have also been taken into account. The effects of the plasma response to externally-applied RMPs have been included in the modelling. The updated results show that computed neutron rates in the presence of RMPs with the plasma response and CX reactions taken into account agree very well with the experimental data throughout the analysis target time. CX reactions play an important role in determining the neutron rates, in particular before the onset of RMPs.

Keywords: fast ions, Monte-Carlo, non-steady-state, neutron rate, MAST

## 1. Introduction

Resonant magnetic perturbations (RMPs) are externally-applied three-dimensional (3D) modifications to axisymmetric tokamak magnetic fields, the purpose of which is to mitigate or suppress edge localized modes (ELMs). They also potentially have the unintended consequence of degrading the confinement of energetic ions, including fusion alpha-particles in the case of a burning plasma, due to the violation of toroidal canonical momentum conservation arising from any non-axisymmetric fields (the orbits of collisionless particles in such configurations are discussed in [1]). It is intended that RMPs will be used to mitigate ELMs in the burning plasma ITER device, and predictive simulations for both the baseline [2,3] and steady-state [4,5] scenarios in ITER indicate that these perturbations could reduce significantly the neutral beam power available to heat the plasma. Modest levels of RMP-induced fusion α-particle losses are also predicted. The possible impact of RMPs on fast ion confinement in ITER motivates us to seek a first principles-based understanding of this impact in existing devices. This is the primary aim of the current paper.

Clear experimental evidence for a degradation of energetic ion confinement resulting from the application of RMPs has been obtained in two conventional tokamaks, ASDEX Upgrade [6] and DIII-D [7]. The application of RMPs to neutral beam-heated pulses in the Mega Amp Spherical Tokamak (MAST) with relatively low plasma current (400kA) was found to be correlated with substantial drops (almost a factor of two) in the measured fluxes of both neutral and charged fusion products [8]. Since almost all the fusion reactions in MAST were either beam-thermal or beam-beam, the drops in neutron and charged fusion product fluxes could be attributed to a loss of beam ion confinement. Simultaneous drops in fast ion deuterium-alpha (FIDA) emission provided further evidence for this interpretation.

The MAST neutron data reported in Ref. [8] have recently been modelled as part of the verification and validation of a newly-developed non-steady-state orbit-following Monte-Carlo (NSS OFMC) code [9]. Specifically, fast ions sampled from a realistic neutral beam deposition profile were tracked from birth to thermalization or loss using field and plasma parameters whose temporal evolution was constrained by the available experimental data in one specific pulse, #30086. Both of the MAST beamlines were used in this pulse, with primary injection energies $E_b$ of 71 keV (beam power $P_{NBI}$ = 2.053MW) and 62keV ($P_{NBI}$ = 1.515MW). Both beamlines were parallel to the plasma current, with tangency radius equal to 0.7m, somewhat inboard of the magnetic axis in this pulse ($\approx$1.04m). The partition of beam power between $E_b$, $E_b/2$ and $E_b/3$ is given for each beamline in Table 2 of Ref. [9]. With these parameters specified, the beam deposition profiles are then determined by the plasma density profile, which is discussed below. The magnetic axis of this pulse lay below the beamlines, with the result that the beam deposition profiles were relatively broad. The fast ion distribution function was used to compute an evolving total neutron rate which could be compared directly with that measured using the MAST fission chamber. The 3D perturbations to the fields were represented by both RMPs and toroidal field (TF) ripples, both calculated in the vacuum approximation. Significant differences were found between simulated neutron rates obtained using full orbit and guiding centre schemes to compute the fast ion orbits, indicating the importance of finite Larmor radius effects in the relatively low magnetic field of MAST (~0.4T). In a separate study of RMP-induced fast ion transport in MAST, carried out using a guiding-centre code [10], relatively modest fast ion losses were observed (a few percent), but this is likely to have been due in part to the fact that this calculation was performed for the case of an up-down symmetric plasma, with a fast ion deposition profile that was more centrally-peaked than that of the pulses discussed in Ref. [8].

Two important effects that were neglected in the modelling reported in Ref. [9] were, first, the plasma response to the applied RMPs, and, second, losses of fast ions due to charge-exchange (CX) reactions with neutral particles. In the case of MAST, it has been found [11] that a toroidally-rotating plasma response typically suppresses resonant components of the RMP field (that is to say, components with toroidal and poloidal mode numbers $m$ and $n$ such that the local safety factor $q = m/n$). However, non-resonant components of the perturbation can be strongly amplified, with the result that the overall effect of the RMPs on fast ion losses could be increased. The amplification of non-resonant components is due to coupling of the external perturbations to marginally stable kink modes with the same $n$ [11-13]. Since low $n$ kink modes tend to have broad eigenfunctions, this coupling process can have the effect of causing greater penetration of RMP fields into the plasma. The loss mechanism may itself be a resonant process, but the resonances in this case would be between the motion of the fast ions and a static field perturbation, which could be either resonant or non-resonant in terms of the local value of $q$. It is known from previous studies that CX losses can have a significant effect on the fast ion distribution in MAST [14], and therefore such losses should be taken into account for the purpose of modelling the total neutron rate.

Recently we have made a number of improvements to NSS OFMC, including the incorporation of CX reactions with background neutrals and subsequent reionization of fast neutrals. Details of the model used to represent these processes are given in section 2. We have also updated the calculation of fusion reaction rates in the code using a parametrization of the cross-section as a function of energy in the centre-of-mass frame given by equations (8-9) in Ref. [15] and the parameters shown in Table IV of that paper. In the process of doing this, we found that the beam-thermal D-D reaction rate given by the previous version of NSS OFMC underestimated the rate by about 50%, due to the $D(d,n)^3He$ cross-section being used instead of the total D-D cross-section. We stress that this underestimate was used only in the version of NSS-OFMC that was developed for the previous MAST analysis, and not in other versions of OFMC. Also, in the old version the birth velocities of NBI fast ions were specified by randomly selecting the ion gyro-phase. It is more realistic to assume that the initial gyro-phase of an NBI fast ion is determined by the beam-line geometry and the magnetic field at the birth point. In this paper we re-analyze the experimental data of MAST pulse #30086 using the updated NSS OFMC code, taking into account CX losses and the plasma response to the RMPs. In the present work, all the calculations were made using the revised fusion reaction rate and with initial velocities determined purely by the beam-line geometry and the local magnetic field.

As part of the updated analysis, the temporal evolution of electron plasma density ($n_e$) and electron temperature ($T_e$) profiles used in the NSS OFMC simulations was also modified to give a better fit to the experimental (Thomson scattering) data of MAST pulse #30086 before, during and after the period in which the RMP coil currents were ramped up (0.35s-0.37s). Specifically, $n_e$ and $T_e$ were assumed to change more slowly after the RMPs were turned on than in the previous analysis, as shown in figure 1 (note that the times corresponding to these profiles have been modified, but the profiles themselves have not changed). The assumed temporal variation of the plasma toroidal rotation was also modified to give a better fit to the measured (CX) data, as shown by the dashed curve in figure 2. In this figure $f_{coil}(t)$ represents the RMP coil current normalised to its maximum value, while $f_{rot-Old}(t)$ and $f_{rot-New}(t)$ are, respectively, previous and updated representations of the plasma rotation rate normalised to its maximum value before the application of the RMPs. We assume that the shape of the rotation profile did not change in time; the model rotation profile before the application of RMPs, which is again based on CX measurements, is that shown in figure 8(C) in Ref. [9]. The other calculation parameters were the same as those listed in Ref. [9]. As before, the ion temperature $T_i$ was assumed to be equal to $T_e$. CX measurements show that $T_i$ was in fact about 10-20% higher than $T_e$ across the plasma and at the times of interest. Changes in $T_i$ have a direct effect on the thermonuclear contribution to the neutron rate, but this has no significant impact on the total neutron emission, since the thermonuclear component of this in MAST was about 1%. A modest underestimate in $T_i$ has some effect on the beam-thermal neutron rate, but this effect is also very small since the primary beam injection speed was about eight times the maximum ion thermal speed [16]. The primary injection energy was sufficiently large compared to $T_e$ that the beam ions slowed down primarily as a result of electron drag.

Simulated neutron rates obtained using the previous and updated versions of NSS OFMC, with and without CX effects taken into account, are compared for the case of vacuum RMPs in section 2. NSS OFMC calculations of neutron rates are compared with those computed with TRANSP in section 3, and rates obtained in the presence of RMPs, with the plasma response and CX reactions taken into account, are shown in section 4. The conclusions drawn from this study are presented in section 5.

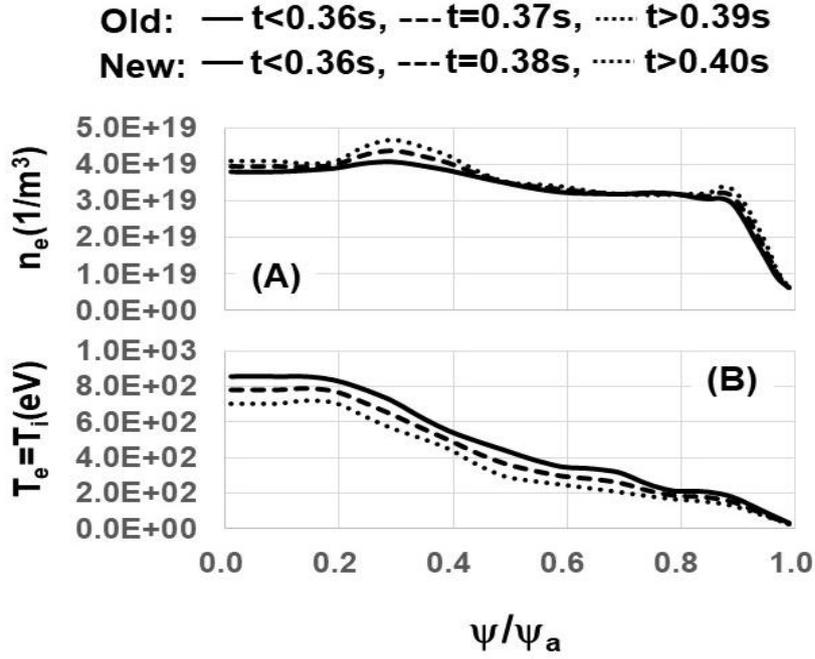

**Figure 1.** Radial profiles of plasma density (A) and temperature (B) at time $t < 0.36$s, $t = 0.37$s and $t > 0.39$s for the previous analysis and $t < 0.36$s, $t = 0.38$s and $t > 0.40$s for the updated analysis.

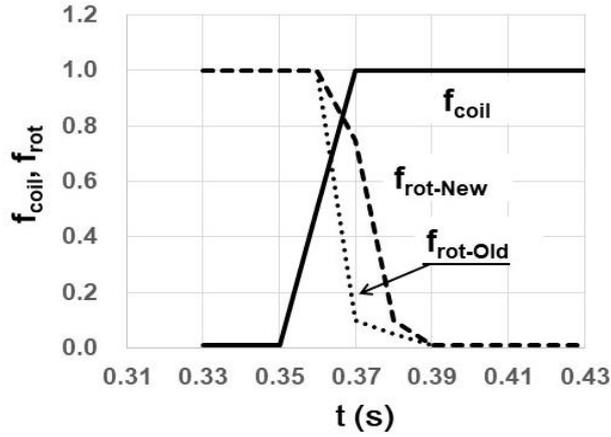

**Figure 2.** Time-variation factors for the RMP coil current $f_{coil}$, and the previous and updated normalised plasma rotation rates $f_{rot\text{-}Old}$, $f_{rot\text{-}New}$ are shown by the solid, dotted and dashed curves respectively.

## 2. Neutron rates from NSS OFMC for vacuum RMPs

For the purpose of comparison with the results presented in Ref. [9], a simulation was first performed using the updated version of NSS OFMC using a full orbit scheme to track the fast ion orbits, with vacuum $n = 3$ RMP and TF ripple fields, and with CX effects neglected. Fast ions were assumed to be lost if they crossed an axisymmetric notional first wall, approximating the true three-dimensional geometry of plasma-facing components in the MAST vacuum vessel (see figure 5 in [9]). Details of the calculation of neutron rates in the non-steady state version of OFMC can be found in Ref. [9]. In brief, the code is used to track the full orbits of fast ions over a time interval which is intermediate between the fast ion orbital period (bounce or transit time) and the collisional slowing-down time. For a given radial slice of the plasma, the appropriate fusion cross-section, together with the fast ion distribution function and the values of bulk plasma parameters, are then used to compute the number of neutrons produced in this time interval. Finally, a summation over radial slices is performed to obtain the total neutron rate as a function of time.

The temporal evolution of the re-analyzed neutron rate is indicated by the closed triangles in figure 3. The experimental (fission chamber) data are shown by the black solid curve in figure 3. There is a substantial difference between the new results and the experimental data, even in the time before the onset of RMPs ($t \leq 0.35$s). This motivates us to consider CX effects as a possible cause of fast ion losses which could occur in an axisymmetric field.

Both CX reactions of fast ions with background neutrals and the subsequent reionization with bulk plasmas were taken into account in the new calculations. The two-dimensional distributions of wall (Franck-Condon), halo and recombination neutral particles in the background plasma were individually calculated using Monte-Carlo techniques [17] and were summed up. The absolute value of the wall neutral density was determined by making the steady-state assumption that the total ionization rate of neutrals was equal to the bulk-plasma particle loss rate defined by the particle confinement time, which was prescribed as an input parameter. For the non-steady state calculations, the 2D neutral profiles were calculated as described above using the plasma parameters and NBI conditions at times $t = 0.36$s and $t = 0.40$s in MAST pulse #30086, assuming a particle confinement time of 10ms. Using the standard definition of the energy confinement time $\tau_E$ as the ratio of plasma stored energy to auxiliary input power, we find that $\tau_E$ ranged from 16ms at $t = 0.36$s to 10ms at $t = 0.40$s. The particle confinement time used to model the 2D neutral profiles is thus reasonable if we assume that it was comparable to $\tau_E$. In the simulations the neutral distribution inferred for $t = 0.36$s was applied to the period prior to this time, while the neutral distribution inferred for $t = 0.40$s was applied to later times. For the period $0.36$s $< t < 0.40$s, the neutral distribution was obtained by linear interpolation between these two profiles. The flux surface-averaged radial distribution of the total neutral density is indicated in figure 4(A) at $t = 0.36$s and $t = 0.40$s by the solid and dashed curves, respectively. For the NBI primary injection energies in MAST pulse #30086 (71keV and 62keV), the CX ionization cross-section accounts for the major part of the total stopping cross-section, and consequently the halo neutrals are dominant in the plasma core region, as shown in Fig.4(A). The ratio of the slowing-down time $\tau_s$ to the CX time $\tau_{CX}$ is a good measure of the importance of fast ion losses due to CX. The CX time is given by

$$\tau_{CX} = \frac{1}{n_0 <\sigma v>_{CX}}, \qquad (1)$$

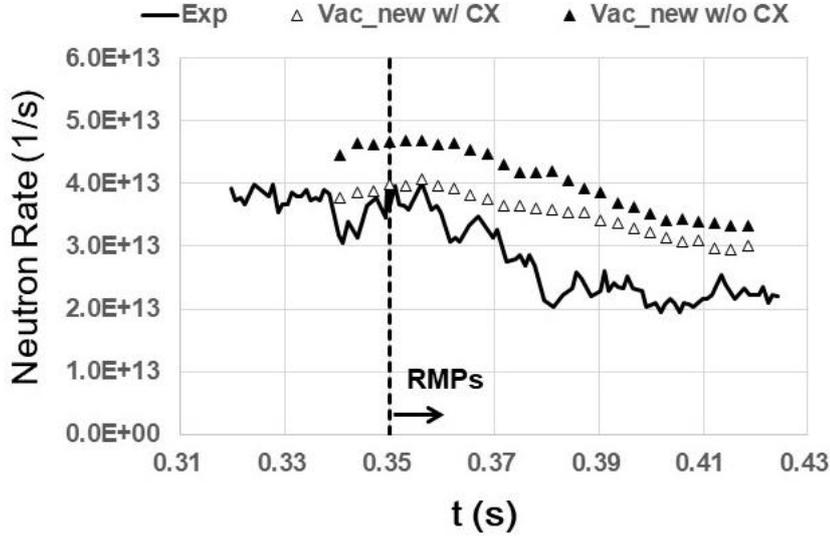

**Figure 3**. Revised neutron rates obtained using full orbit tracking in NSS OFMC for MAST shot #30086 with vacuum RMPs are shown as closed triangles. The neutron rates obtained with vacuum RMPs and CX reactions are represented by open triangles. The solid curve shows the experimental (fission chamber) data.[1]

---

[1] As shown in Ref. [9] (see open circles in figure 9 of this paper), the neutron rate computed previously for this pulse using NSS OFMC with the same model field perturbations (vacuum RMPs and TF ripple) dropped from $3.4\times10^{13}$s$^{-1}$ before the onset of RMPs ($t < 0.35$s) to $2.6\times10^{13}$s$^{-1}$ at $t > 0.39$s. Generally, the updated neutron rates are about 30% higher than the previously-computed values.

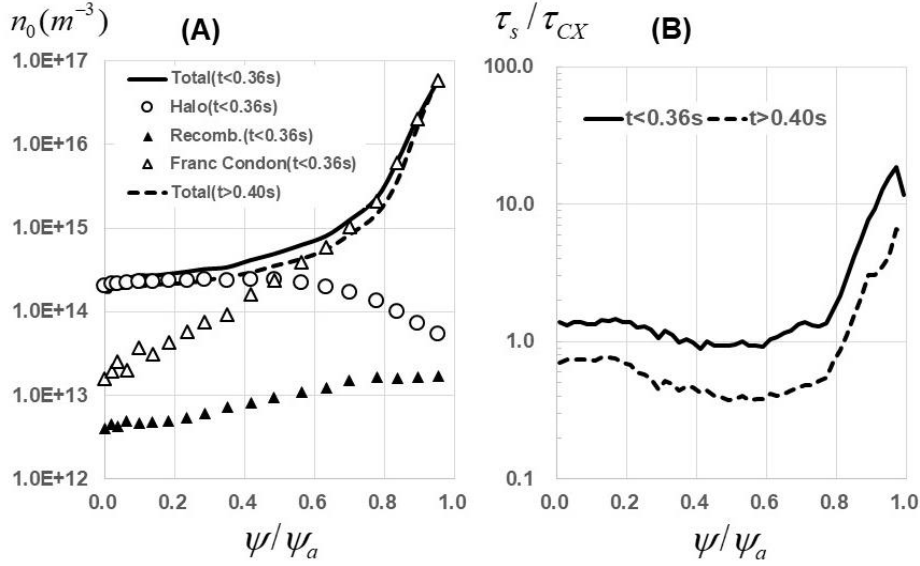

**Figure 4.** Flux surface-averaged radial distribution of the total neutrals (A) at times $t \leq 0.36$s and at $t \geq 0.40$s are shown by the solid and dashed curves, respectively. The contributions of different neutral components [halo, recombination and wall (Franck-Condon) neutrals] at $t \leq 0.36$s are also shown in (A). The surface-averaged ratios of the slowing down time to the CX time $\tau_s/\tau_{CX}$ for the main beam energy $E_B = 71$keV at times $t = 0.36$s and 0.40s are indicated in (B) by the solid and dashed curves, respectively.

where $n_0$ is the neutral density and $<\sigma v>_{CX}$ the flux surface-averaged CX reaction rate. Flux surface-averaged values of $\tau_s/\tau_{CX}$ for a primary beam injection energy $E_B = 71$keV at times $t \leq 0.36$s and $t \geq 0.40$s are indicated in figure 4(B) by the solid and dashed curves, respectively.

As shown in figure 4, the neutral density in the core region, which is mainly due to halo particles, is higher than $10^{14}$m$^{-3}$, causing the ratio $\tau_s/\tau_{CX}$ to be as high as 1.0 even in this part of the plasma, indicating that a substantial fraction of NBI fast ions are likely to have been lost due to CX before they had slowed down in this pulse.

The simulated neutron rates with CX reactions taken into account are shown in figure 3 by the open triangles. These agree very well with the experimental data before the onset of RMPs ($t < 0.35$s). However, there is still a very large gap between the neutron rates in this simulation and the MAST data after the onset of RMPs. The effect of CX loss of fast ions on the neutron rates is very significant before the onset of RMPs. The effect, however, is relatively small after the onset of RMPs. This implies that the loss time of fast ions due to RMPs is shorter than the CX loss time.

## 3. Comparison of neutron rates from NSS OFMC and TRANSP

The neutron rates obtained using the NSS OFMC code were compared with those computed with the NUBEAM fast particle module of the TRANSP code [18]. For this comparison, calculations using NSS OFMC were made for an axisymmetric field and using a guiding-centre orbit-following scheme to track the fast ions (NUBEAM also uses guiding-centre tracking). The results with and without CX reactions are shown in figure 5 by open and closed squares, respectively. The results from TRANSP without anomalous (non-classical) diffusion are shown by the solid curve, taken from figure 4 in Ref. [8]. Like the updated version of NSS OFMC, TRANSP/NUBEAM models the interaction of beam ions with neutrals, but the neutral density used to generate the TRANSP results shown in figure 5 was about an order of magnitude lower than that used in the NSS OFMC simulations. As noted in section 2, the NSS OFMC neutral density profiles correspond to a particle confinement time that is comparable to the known energy confinement time during the relevant time interval in this pulse, and these profiles are thus likely to be closer to the true profiles than those used in the TRANSP simulation.

The open and closed squares in figure 5 are close together, indicating that in this case CX reactions have only a small effect on the predicted neutron rate. Both rates are also in good agreement with those from TRANSP at late times, but the TRANSP rate is about 10% lower at early times ($t < 0.37s$). The source of this small discrepancy is not known, but it could be due to minor differences in the evolving bulk plasma profiles used in the modelling. For reference, results using the full orbit-following scheme in NSS OFMC are also shown in figure 5 by the crosses. As shown in figure 11(B) in Ref. [9], the neutron yield $Y_n$ changes nonlinearly along the full orbit with respect to the radial position. As a result, the gyro-averaged neutron yield in the full orbit calculation is smaller than that obtained by tracking the particle guiding centre. The local value of the slowing-down time $\tau_s$ changes in a similar fashion along the full orbit. The slowing-down time and the neutron rate averaged over one bounce motion of the guiding centre were calculated for NBI-produced fast ions with an injection energy 71keV using both full-orbit and guiding-centre following schemes. The ensemble-averaged slowing-down time and the neutron yield per fast ion calculated in the full-orbit following scheme, $<\tau_{s-FO}>$ and $<Y_{n-FO}>$, were compared to those in the guiding-centre following scheme, $<\tau_{s-GC}>$ and $<Y_{n-GC}>$. The calculation results are $<\tau_{s-FO}>/<\tau_{s-GC}>= 0.87$ and $<Y_{n-FO}>/<Y_{n-GC}>= 0.97$. The total neutron rate is approximately proportional to the fast ion density, and hence (under steady-state conditions) to the product of the slowing-down time and the yield per fast ion. The difference between the open squares and crosses in figure 5 can thus be attributed mainly to finite Larmor radius effects on the orbit- and ensemble-averaged slowing-down time.

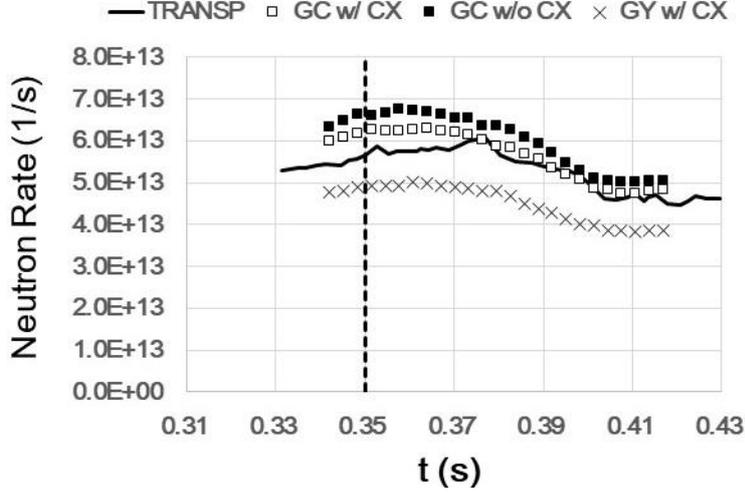

**Figure 5.** Neutron rates obtained from NSS OFMC with guiding-centre (GC) tracking of fast ion orbits for MAST pulse #30086 in an axisymmetric field with and without CX reactions are shown by the open and closed squares, respectively. The results from TRANSP without artificial diffusion are shown by the solid curve. NSS OFMC results obtained using a full-orbit (GY) following scheme are also shown by the crosses. The vertical dashed line indicates the time at which the RMP coil current started to be ramped up in the experiment.

**4. Neutron rates in the presence of RMPs with plasma response**

As shown in figure 3, a large difference remains between the neutron rates calculated with vacuum RMPs and the experimental data in the period after the onset of the RMPs. However, as discussed in section 1, it is well-known that RMPs are substantially modified by the plasma [11-13]. The plasma response to RMPs applied in MAST pulse #30086 was estimated using the MARS-F code, which employs a linear, fully toroidal, single-fluid resistive magnetohydrodynamic (MHD) model [19]. Due to aliasing effects, RMPs with dominant toroidal mode number $n_{\mathrm{rmp}}$ in a tokamak equipped with $n_{\mathrm{coils}}$ RMP coils have a significant sideband harmonic with mode number $n_{\mathrm{coils}} - n_{\mathrm{rmp}}$ [20]. In the case of MAST $n_{\mathrm{coils}} = 12$ and, as mentioned previously, the RMPs in pulse #30086 had $n_{\mathrm{rmp}} = 3$, so that the total field perturbation had a large amplitude sideband with $n = 9$. This is taken into account in the MARS-F modelling of the plasma response to the RMPs in #30086. Figure 6 shows the computed poloidal dependence of the radial component $\delta_R$ of (A) plasma response RMPs, (B) vacuum RMPs and (C) TF ripple along magnetic surfaces with normalized poloidal flux $\psi = 0.6$ (solid curves), $\psi = 0.8$ (dashed curves) and $\psi = 1.0$ (dotted curves). The quantity $\delta_R$ is defined here as $(B_{R\text{-max}} - B_{R\text{-min}})/(2B_{ax})$ where $B_{ax} \approx$ 0.39T is the vacuum magnetic field at the magnetic axis and $B_{R\text{-max}}$, $B_{R\text{-min}}$ are maximum and minimum values in the toroidal direction. An important point to note from this plot is that while the plasma has very little effect on the maximum field perturbation, the plasma response RMP field has a significant amplitude over a much larger range of poloidal angles than the vacuum RMP field, which is confined to a region close to the RMP coils. This illustrates the point made in section 1 that the coupling of RMPs with plasma eigenmodes can cause greater penetration of the perturbation into the plasma. For values of $\psi$ closer to the magnetic axis than those used in figure 6 show, we find that $\delta_R$ is negligibly small for all poloidal angles in the vacuum case, but still significant in the plasma response case. Given that the fast ion density falls off rapidly with distance from the magnetic axis, it may be expected that the effects of the plasma response on the confinement of those ions could be substantial.

The neutron rates in the presence of RMPs that include the effects of the plasma response, with and without CX reactions, are indicated in figure 7 by open and closed circles, respectively. For comparison, the neutron rates in the presence of vacuum RMPs with and without CX reactions, which appear in figure 3, are also shown in figure 7 by open and closed triangles, respectively. For reference, the neutron rates in an axisymmetric field without CX reactions are shown by the crosses. All the results shown in figure 7 were obtained using the full-orbit following scheme in NSS OFMC.

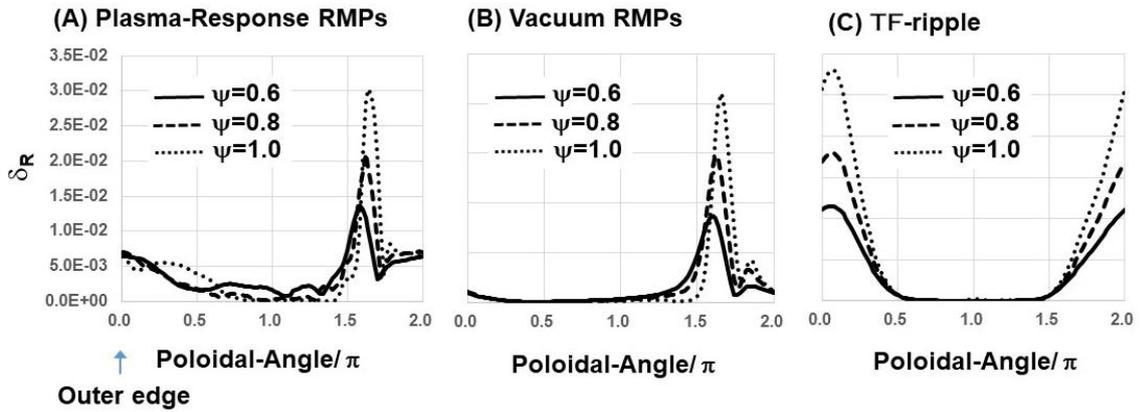

**Figure 6.** Poloidal variation of the radial component of plasma response RMPs (A), vacuum RMPs (B) and TF ripple (C) along the magnetic surfaces with normalized poloidal flux $\psi = 0.6$ (solid curves), $\psi = 0.8$ (dashed curves) and $\psi = 1.0$ (dotted curves). $\delta_R$ is defined as $(B_{R\text{-max}} - B_{R\text{-min}})/(2B_{ax})$ where $B_{ax} \approx 0.39$T is the vacuum magnetic field at the magnetic axis and $B_{R\text{-max}}$, $B_{R\text{-min}}$ are maximum and minimum values in the toroidal direction.

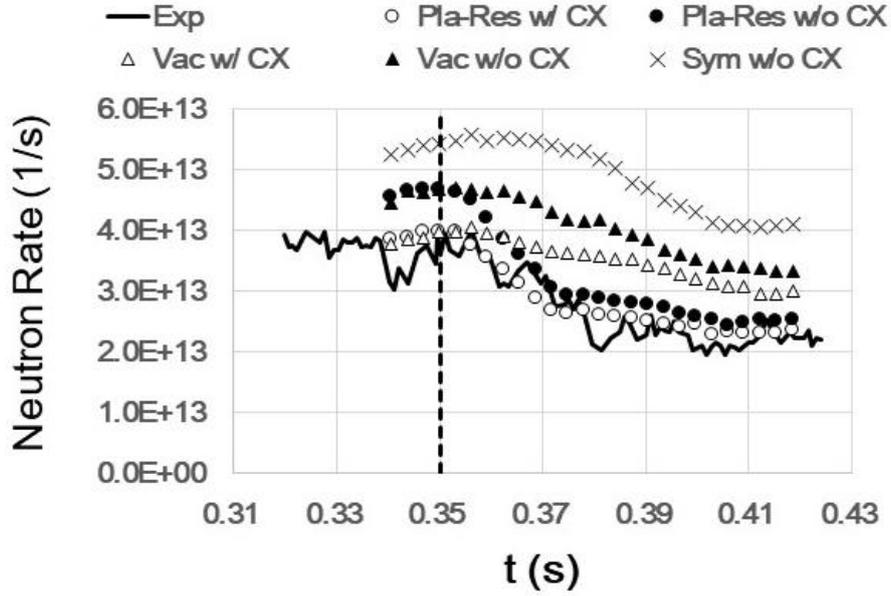

**Figure 7.** Revised neutron rates obtained using full orbit tracking in NSS OFMC for MAST pulse #30086, taking into account the plasma response to RMPs, are shown by open and closed circles for simulations with and without CX reactions, respectively. The revised neutron rates for vacuum RMPs, with and without CX reactions, are shown by open and closed triangles, respectively. The experimental (fission chamber) data are shown by the solid black curve. The neutron rates in an axisymmetric field with CX reactions are shown by the crosses.

The results shown by the crosses in figure 7 indicate that even the relatively small changes in plasma parameters shown in figure 1 can cause a substantial reduction in neutron rates. This implies that a non-steady-state treatment, such as that employed in Ref. [9] and in the present paper, is required for the quantitative analysis of RMP-induced fast ion losses in pulses such as this. The difference between the crosses and the closed triangles before the onset of RMPs (about 15%) is due to the effects of TF ripple, while the difference after the onset of RMPs is due to the combined effects of TF ripple and RMPs. The reduction of the neutron yield due to TF ripple alone is thus quite significant in pulse #30086. As discussed in Ref. [9], this is likely to be due to the relatively low plasma current in this particular pulse, causing radial excursions of fast ions into regions with high ripple amplitude. As shown in figure 7, the neutron rates in the presence of RMPs with both the plasma response and CX reactions taken into account agree very well with the experimental data throughout the analysis target time. Comparing results obtained with and without the plasma response to RMPs, we conclude that it is essential to include this response in the modelling in order to reproduce accurately the experimental data from MAST.

## 5. Summary

A number of improvements have recently been implemented in the NSS OFMC code, in particular the inclusion of a scheme for representing charge-exchange losses of fast ions in addition to those arising from RMPs. The parametrization of fusion reaction rates in the code has also been updated. As an application of the code, neutron rates in a MAST pulse with RMPs have been recalculated, taking into account not only the loss of NBI fast ions in 3D fields representing RMPs (with the plasma response to these perturbations taken into account) and toroidal field ripple, but also the loss of fast ions via CX reactions with background neutrals and the subsequent reionization with bulk plasma. The reanalyzed neutron rates agree very well with the experimental data throughout the analysis target time. In order to reproduce the measured neutron rates, it is essential to model the plasma response to the RMPs, and CX reactions also play an important role, in particular before the onset of RMPs. After the onset of RMPs, CX effects are less important, possibly due to the loss time associated with the RMPs being shorter than that due to CX.

It should be noted that the NSS OFMC modelling reported here does not include all of the possible processes that could have contributed to the fast ion transport in MAST pulse #30086. In particular the effects of MHD instabilities and ELMs were not taken into account. However, as noted in Ref. [8], both MHD activity and ELMs were of relatively low amplitude during the period of RMP application in this pulse, and it is therefore likely that they made only a small contribution to the fast ion losses. In any case the results presented in this paper provide useful insights into the relative importance of RMPs, the plasma response to such perturbations, TF ripple and charge exchange reactions with neutrals in determining the overall fusion performance of spherical tokamak plasmas.

## Acknowledgement

The authors gratefully acknowledge Mr. M. Suzuki and Dr. J. Shiraishi at the National Institutes for Quantum and Radiological Science and Technology for their technical support in updating the NSS OFMC code. This work was funded in part by the RCUK Energy Programme [under grant EP/P012450/1]. To obtain further information on the data and models underlying this paper please contact PublicationsManager@ccfe.ac.uk.

## References

[1] Helander P 2014 *Rep. Prog. Phys.* **77** 087001

[2] Shinohara K, Tani K, Oikawa T, Putvinski S, Schaffer M and Loarte A 2012 *Nucl. Fusion* **52** 094008

[3] Varje J, Asunta O, Cavinato M, Gagliardi M, Hirvijoki E, Koskela T, Kurki-Suonio T, Liu Y Q, Parail V, Saibene G, Sipilä S, Snicker A, Särkimäki K and Äkäslompolo S 2016 *Nucl. Fusion* **56** 046014

[4] Shinohara K, Kurki-Suonio T, Spong D, Asunta O, Tani K, Strumberger E, Briguglio S, Koskela T, Vlad G, Günter S, Kramer G, Putvinski S, Hamamatsu K and ITPA Topical Group on Energetic Particles 2011 *Nucl. Fusion* **51** 063028

[5] Tani K, Shinohara K, Oikawa T, Tsutsui H, Miyamoto S, Kusama Y and Sugie T 2012 *Nucl. Fusion* **52** 013012

[6] Garcia-Munoz M *et al.* 2013 *Plasma Phys. Control. Fusion* **55** 124014

[7] Van Zeeland M *et al.* 2015 *Nucl. Fusion* **55** 073028

[8] McClements K G, Akers R J, Boeglin W U, Cecconello M, Keeling D, Jones O M, Kirk A, Klimek I, Perez R V, Shinohara K and Tani K 2015 *Plasma Phys. Control. Fusion* **57** 075003

[9] Tani K, Shinohara K, Oikawa T, Tsutsui H, McClements K G, Akers R J, Liu Y Q, Suzuki M, Ide S, Kusama Y and Tsuji-Iio S 2016 *Plasma Phys. Control. Fusion* **58** 105005

[10] Pfefferlé D, Misev C, Cooper W A and Graves J P 2015 *Nucl. Fusion* **55** 012001

[11] Liu Y Q, Kirk A, Sun Y, Cahyna P, Chapman I T, Denner P, Fishpool G, Garofalo A M, Harrison J R, Nardon E and the MAST team 2012 *Plasma Phys. Control. Fusion* **54** 124013

[12] Turnbull A D, Ferraro N M, Izzo V A, Lazarus E A, Park J-K, Cooper W A, Hirshman S P, Lao L L, Lanctot M J, Lazerson S, Liu Y Q, Reiman A and Turco F 2013 *Phys. Plasmas* **20** 056114

[13] Bird T M and Hegna C C 2014 *Phys. Plasmas* **21** 100702

[14] McClements K G and Thyagaraja A 2006 *Phys. Plasmas* **13** 042503

[15] Bosch H -S and Hale G M 1992 *Nucl. Fusion* **32** 611

[16] Klimek I, Cecconello M, Gorelenkova M, Keeling D, Meakins A, Jones O, Akers R, Lupelli I, Turnyanskiy M, Ericsson G and the MAST Team 2015 *Nucl. Fusion* **55** 023003

[17] Shimizu K, Takizuka T, Sakurai S, Tamai H, Takenaga H, Kubo H and Miura Y 2003 *J. Nucl. Mater.* **313-316** 1277

[18] Pankin A, McCune D, Andre R, Bateman G and Kritz A 2004 *Computer Phys. Commun.* **159** 157

[19] Liu Y Q, Kirk A, Gribov Y, Gryaznevich M P, Hender T C and Nardon E 2011 *Nucl. Fusion* **51** 083002

[20] Ham C J, Cramp R G J, Gibson S, Lazerson S A, Chapman I T and Kirk A 2016 *Nucl. Fusion* **56** 086005